\documentclass[aps,prl,reprint,groupedaddress]{revtex4-1}

\usepackage{graphicx}
\usepackage{amsmath} 

\begin{document}


\title{Imaging the Conductance of Integer and Fractional Quantum Hall Edge States}


\author{Nikola Pascher$^{1}$}
\email[]{npascher@phys.ethz.ch}
\author{Clemens R\"ossler$^{1}$}
\author{Thomas Ihn$^{1}$}
\author{Klaus Ensslin$^{1}$}
\author{Christian Reichl$^{1}$}
\author{Werner Wegscheider$^{1}$}
\affiliation{Solid State Physics Laboratory, ETH Zurich, 8093 Zurich, Switzerland}


\date{\today}

\begin{abstract}
We measure the conductance of a quantum point contact (QPC) while the biased tip of a scanning probe microscope induces a depleted region in the electron gas underneath. At finite magnetic field we find plateaus in the real-space maps of the conductance as a function of tip position at integer ($\nu=1,2,3,4,6,8$) and fractional ($\nu=1/3,2/3,5/3,4/5$) values of transmission. They resemble theoretically predicted compressible and incompressible stripes of quantum Hall edge states. The scanning tip allows us to shift the constriction limiting the conductance in real space over distances of many microns. The resulting stripes of integer and fractional filling factors are rugged on the micron scale, i.e. on a scale much smaller than the zero-field elastic mean free path of the electrons. Our experiments demonstrate that microscopic inhomogeneities are relevant even in high-quality samples and lead to locally strongly fluctuating widths of incompressible regions even down to their complete suppression for certain tip positions. The macroscopic quantization of the Hall resistance measured experimentally in a non-local contact configuration survives in the presence of these inhomogeneities, and the relevant local energy scale for the $\nu=2$ state turns out to be independent of tip position.
\end{abstract}


\maketitle

\section{Introduction}
Two-dimensional electron gases at high magnetic fields applied normal to the electron gas plane exhibit the quantum Hall effect. The integer \cite{Klitzing1980} and fractional \cite{Tsui1982} quantum Hall effects are two different macroscopic quantum phenomena, which lead to surprisingly similar observations in electron transport experiments. The similar phenomenology of both effects is commonly believed to be a result of one-dimensional channels at the sample edges, which transmit electrical signals without losses between contacts separated by macroscopic distances \cite{Halperin1982,Buttiker1988,Chklovskii1992}.

In the integer quantum Hall effect, each occupied bulk Landau level gives rise to a pair of counterpropagating channels at opposite sample edges \cite{Halperin1982,Buttiker1988}. Each counterpropagating pair contributes one conductance quantum $e^2/h$ to the total conductance of the quantum Hall fluid, giving a total quantized conductance (Hall conductance) of $e^2/h\times \nu$, where the integer filling factor $\nu$ is the number of Landau levels occupied in the bulk \cite{Buttiker1988,Beenakker1991}. Spin degeneracy of edge channels can be lifted by Zeeman splitting enhanced by exchange interaction effects leading to a spatial substructure \cite{Dempsey1993,Ihnatsenka2006}. Edge channels in the fractional quantum Hall effect have been theoretically proposed \cite{Weiss2011,Chang1990,MacDonald1990,Wen1990,Gelfand1994}. Exchange and correlation effects can bring about spatial substructure beyond the spin-splitting of Landau levels, an effect called edge reconstruction \cite{Chamon1994,Wan2002,Yang2003}.  Concerning transport, however, theory predicts the same form of the quantized conductance as in the integer regime, but with fractional values of $\nu$, in agreement with the experiment \cite{Tsui1982}.

Early experiments showed spatially resolved edge channel transport with various techniques \cite{Haren1995,Knott1995,Wei1998}. Alternative local investigations of the integer quantum Hall effect employed scanning probe techniques, such as scanned potential microscopy \cite{McCormick1999}, the scanning SET \cite{Yacoby1999}, or subsurface charge accumulation imaging \cite{Finkelstein2000}. Local gate electrodes enable selective backscattering experiments of integer \cite{Haug1993} or fractional \cite{Chang1989,Kouwenhoven1990} quantum Hall edge channels. Similarly, narrow quantum point contact constrictions allow experimentalists to control local scattering between counterpropagating integer \cite{vanWees1989,Snell1989} or fractional edge channels \cite{Milliken1996,Glattli2000,Roddaro2003}. 
Scanning gate microscopy (SGM) experiments on quantum point contacts in the integer quantum Hall regime have complemented the conventional transport experiments by inducing a local potential perturbation in the electron gas near the quantum point contact with the scanning tip \cite{Aoki2005,Paradiso2012}. This arrangement leads to tip-controlled spatially resolved selective backscattering of integer edge channels. Most recently, an experiment by Paradiso and coworkers \cite{Paradiso2012} focused on exploring the physics of fractional quantum Hall edge states. The experiment found evidence for the transmission and backscattering of fractional edge channels through a constriction by a statistical analysis of scanning gate images, taken at a temperature of 250~mK.

In this paper we report scanning gate experiments at an electron-temperature of 170~mK, which explore the formation of integer and fractional quantum Hall edge channels in a constriction under the influence of a scanning tip. Fulfilling the hopes raised by Ref. 14, our measurements exhibit fractional transmission in the raw data without a statistical analysis. We find unprecedented rich structure on the local scale, which we interpret as an interplay between the interaction-driven local formation of correlated states leading to edge reconstruction, and small residual potential variations occurring in our high-mobility sample on typical length scales of a micrometer, far below the zero-field elastic mean free path of electrons.

\begin{figure}
\includegraphics[width=\columnwidth]{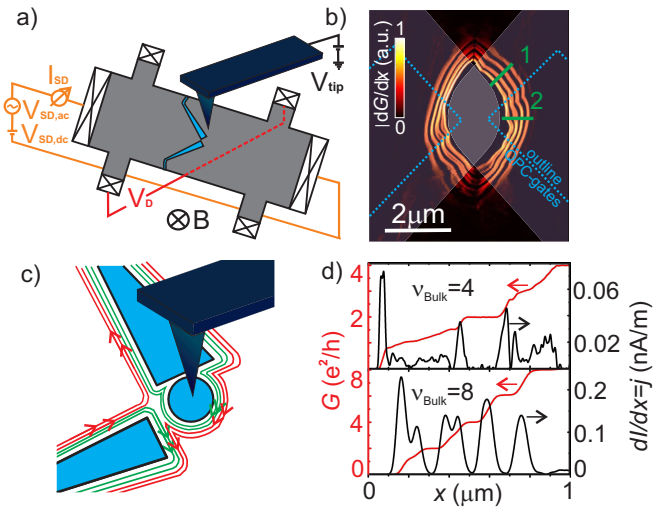}
\caption{a) Schematics of the experimental setup. All indicated parameters are recorded as a function of tip position. The QPC-top gates are highlighted in blue. The QPC forms at the minimum separation of the two gates, which is 800 nm. b) Typical differential conductance map measured at a bulk filling factor of $\nu_{bulk}$=8, corresponding to an external magnetic field of 1 T. The general shape of the observed feature can be understood as an electrostatic fingerprint of the depletion zones of the gates, as highlighted with the white-shaded areas. c) Blow-up of the region around the QPC in Fig. 1a. The tip is used to selectively backscatter edge channels. In the sketch shown here, two spin-degenerate edge channels are reflected and two are transmitted, thus the registered conductance will be 2e$^2$/h. d)  Linecuts at the position of the green line number 2 for filling factors $\nu_{Bulk}$=4 and 8 and their numerical derivative with respect to tip position. The conductance changes in steps of e$^2$/h as a function of tip position, depending on how many edge channels get transmitted or reflected. Even filling factors appear as distinct plateaus, odd filling factors as shoulders.}
\label{figure1}
\end{figure}


\section{Observation of quantum Hall states}
Nanostructures were patterned on top of a high-mobility GaAs-AlGaAs-heterostructure wafer with a two-dimensional electron gas (2DEG) forming 120 nm below the surface. The 2DEG has a density of $1.9 \times 10^{11}$~cm$^{-2}$ and a mobility of $3.5 \times 10^6~$cm$^2$/Vs. The 2DEG was etched in the shape of a Hall-bar and Ohmic contacts were deposited. All quantities which are schematically summarized in Fig. 1a) were measured as a function of tip position. The results are obtained with a QPC, where the minimal distance between the two gates is 800 nm. During the experiments the QPC gates were biased at a voltage of -0.5 V, which is enough to deplete the underlying electron gas and to define a narrow channel \cite{Rossler2011}. An ac bias-voltage of $V_{SD,ac}$=20 $\mu$V was applied and the modulated source-drain current $I_SD$ was measured simultaneously with standard lock-in techniques (see Fig. 1a)). This bias-voltage was chosen experimentally such that features did not get broadened. By simultaneously measuring the source-drain current $I_{SD}$ and the diagonal voltage $V_D$ we determine $G=\frac{I_{SD}}{V_D}=\frac{e^2}{h}\nu_{QPC}$ and therefore the number of transmitted QPC-modes $\nu_{QPC}$ \cite{Beenakker1991}. The measurements were carried out at an electron-temperature of 170 mK in a quantizing magnetic field $B$ perpendicular to the 2DEG plane. \\

In the SGM experiment the voltage-biased tip of the microscope was used to induce a local potential perturbation in the 2DEG \cite{Gildemeister2007}. A constant voltage of $V_{tip}$= -4.5 V was applied to the tip, so that the electron gas underneath it gets totally depleted, leading to a disc-shaped region of zero electron density with a radius of about 1.2 $\mu$m. The tip consists of a focused ion beam sharpened platinum-iridium wire (radius ca. 20 nm). \\

In Fig. 1b) we show a typical SGM-picture, measured at a bulk filling factor $\nu_{bulk}$=8, corresponding to a magnetic field $B$=1 T. The colors encode the values of the derivative of the conductance with respect to the $x$ direction. The blue dashed lines show the outline of the metallic gate fingers defining the QPC. The general shape of the rich pattern is mainly caused by electrostatics. We indicate a fingerprint of the shape of the depletion-zones along the gate edges, as shown by the gray shading. The width between the central lens and the outer contour of the bright region is roughly given by the width of the QPC.
The fine-structure deserves a more detailed discussion. 
As illustrated in Fig. 1c), applying a sufficiently negative voltage to the tip leads to a disc-shaped region of total depletion underneath it. When this tip-depleted region comes close to the QPC-gap it will consecutively reflect edge channels (see Fig. 1c)), until it leads to total depletion of the electron gas in the narrow QPC-channel \cite{Paradiso2010}. 
With the filling factor in the bulk being adjusted to 8, as it is the case for Fig. 1b), not accounting for spin-splitting, four spin-degenerate even-integer edge channels will be present. Once the tip approaches the QPC, the two depletion regions of the tip and the negatively biased gates are so close to each other that the innermost edge channels get reflected and the outer edge channels are still transmitted. Consequently the detected QPC conductance $G=\frac{e^2}{h}\nu_{QPC}$ is decreased in steps.
Figure 1d) shows linecuts at the position of the green line number 2 in Fig. 1b) for filling factors $\nu_{bulk}$=4 and 8 and its numerical derivative with respect to tip position. If the tip is far away from the QPC constriction, we measure the bulk filling factor. If it approaches the QPC gap, edge channels get reflected one after the other, leading to stepwise reduced conductance values. The even filling factors are seen as very pronounced plateaus, the odd filling factors are less clearly seen. This is a manifestation of the relative magnitude of the energy gaps dominating transport: Because the separation of Landau levels $\hbar\omega_c$ is by orders of magnitude larger than the energy scale for spin-splitting, the Zeeman-gap $g\mu_BB$ the respective plateaus are more or less pronounced, respectively. 
Plateaus in the conductance as a function of tip position correspond to the length scale where the system can compensate for a change in the electrostatic potential without changing the filling factor in the QPC. Thus one might assume that the width of the plateaus as measured with SGM is related to the width of the incompressible region in the center of the QPC \cite{Suddards2012,Weiss2011,Chklovskii1992,Venkatachalam2011,Aoki2005,Paradiso2012}. Roughly speaking, the first derivative of the measured source-drain current with respect to the tip position corresponds to the current density which flows in the respective edge channel $dI/dx=j(x)$. The images show that the current density is high in-between the plateaus and zero at the position of the plateaus which can be identified as the incompressible stripes. This agrees with the notion that in the quantum Hall regime the transport current flows only in the compressible stripes \cite{Chklovskii1995,Beenakker1990}. At higher magnetic fields significantly more fine structure of compressible and incompressible stripes can be resolved (Fig. 1d)). The current density is especially high at the edge of the compressible stripe. The complicated structure inside the compressible stripe at a bulk filling factor of 4 might support the idea that the edge channels are reconstructed to consist of a fine structure of different integer and fractional contributions \cite{Beenakker1990}. \\

\begin{figure}
\includegraphics[width=\columnwidth]{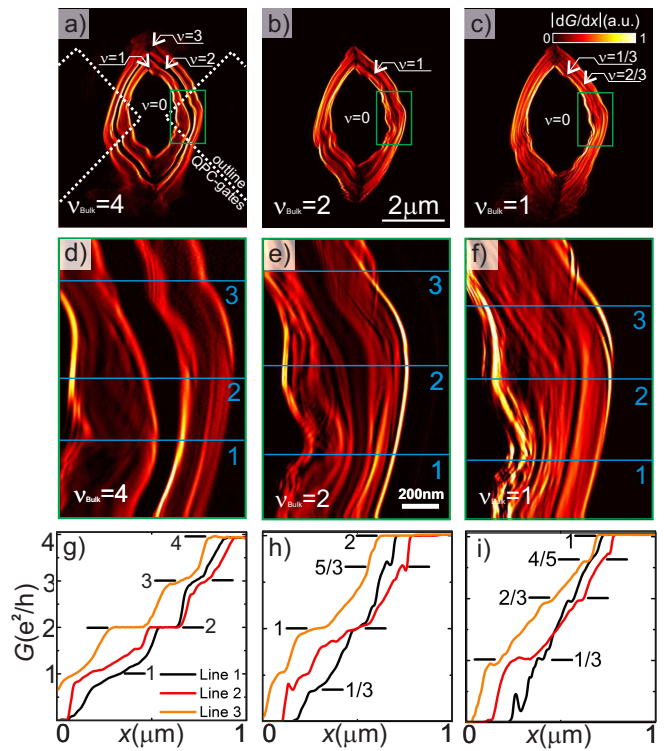}
\caption{a) b) c) Conductance maps at bulk filling factors 4, 2 and 1, corresponding to magnetic field values of 2 T, 3.5 T and 8 T. Even filling factors can be seen as very pronounced rings of constant conductance. Odd filling factors show up as shoulders. d), e) and f) High resolution zooms taken at the position of the green frames in Figs. a) b) c). d) Bulk filling factor of 4. The $\nu$=2 plateau is clearly resolved, $\nu=1$ and $\nu=3$ appear as shoulders. e) Bulk filling factor of 2. f) Bulk filling factor of 1. g), h) and i) show linecuts at the positions of the blue lines. Even, odd and fractional filling factors can be resolved.}
\label{figure2}
\end{figure}

Figure 2 shows a series of images taken at magnetic field values of 2 T, 3.5 T and 8 T, corresponding to bulk filling factors of 4, 2 and 1. In all images the plateaus corresponding to even integer filling factors show up as pronounced dark rings. Odd integer filling factors can be observed as thinner rings, as the corresponding plateaus are less pronounced. The maximum number of resolvable integer plateaus corresponds to the filling factor which was adjusted in the bulk. The higher the magnetic field, the more fine structure can be resolved between integer filling factors. At a bulk filling factor of 2 a pronounced ring with the spin-polarized filling factor 1 can be measured. At a magnetic field of 8 T, where the bulk filling factor is 1 the fine structure is very rich. There are two plateaus of constant fractional filling factor $\nu=1/3$ and $2/3$ going all around the lens.\\

In order to characterize the fine structure between integer plateaus, Figs. 2 d), e) and f) show zooms with high spatial resolution at the position of the green frames in Figs. 2a-c. With the help of the linecuts (Fig. 2g), h), i)) at the positions of the blue lines most of the fine features can be assigned to quantum-Hall plateaus. Others might more likely be caused by disorder-induced antidot- or quantum dot resonances \cite{Paradiso2012/2,Martins2013}. At a bulk filling factor of 4 (Fig. 2d), g)) even and odd filling factors show up as clear plateaus. For a bulk filling factor of 2 (Fig. 2e), h)) we see a clear stripe for the $\nu$=1 plateau. Again the structure is very rich. More features show up at fractional values of e$^2$/h, e.g. 1/3 and 5/3. When the filling factor in the bulk is 1 (Fig. 2f), i)) we see very clear indications for fractional states. There are some regions, where the conductance displays a clear plateau. This allows us to resolve the fractional states with conductance 1/3, 2/3 and 4/5 which can also be observed in conventional transport experiments \cite{Willett1987}.
These findings show that by placing the tip at different positions, the potential landscape of the QPC can be modified in a way that it is more favorable to transmit fractional edge channels. If this is the case, plateaus with fractional filling factors are resolved. 
From these images we can draw conclusions about the edge and its roughness. Along the edge the structure is irregular on a length-scale of a few hundred nanometers, indicating edge roughness on this scale. The conductance maps also show a lot of fine structure in-between the plateaus. This is an indication that conductance at the edge is provided by a fine fabric of different conductance paths \cite{Beenakker1990}.

\begin{figure}
\includegraphics[width=\columnwidth]{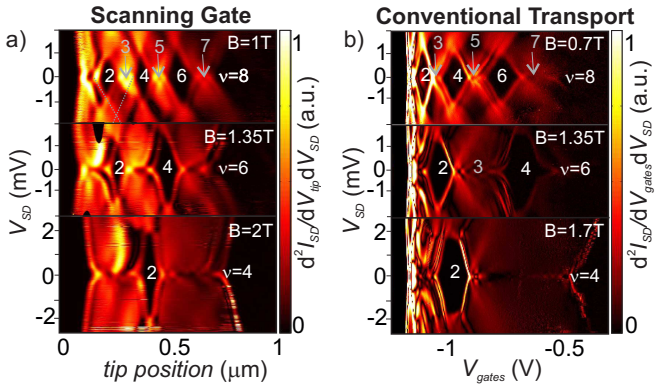}
\caption{False color plots of the transconductance $dI^2_{SD}/dV_{Tip}dV_{SD}$ or $dI^2_{SD}/dV_{Gates}dV_{SD}$. The derivative with respect to the tip is measured directly by putting a small ac modulation on the tip. The derivative with respect to the gates voltage is calculated numerically. Regions of even filling factors appear as black diamonds centered around V$_{SD}$=0mV. The dashed lines in a) show exemplarily how we determined the size of the energy gap. The two data sets are taking at different cool-downs. }
\label{figure3}
\end{figure}

\section{Position independence of the $\nu=2$ energy gap}
SGM offers the possibility to investigate the quantum Hall energy gaps with spatial resolution by applying a finite dc-source drain bias voltage $V_{SD,dc}$. For these measurements a small ac modulation of 50$\mu$V was applied to the tip, so that the transconductance with respect to the tip can be measured. Then, line number 1 in Fig 1a) is repeatedly scanned, while after each line $V_{SD,dc}$ is changed. The derivative with respect to tip position (Fig. 3a)) or gate voltage (Fig. 3b)) is calculated numerically. The results are depicted in Fig. 3a), together with similar transconductance results obtained in conventional transport experiments (Fig. 3b)). Different magnetic field values were applied in a) and b) to compensate for a slightly changed electron density in two different cooldown cycles. The very broad even filling factor plateaus give rise to very clear diamonds, odd filling factors are less pronounced, fractional states can not clearly be resolved in these measurements \cite{Rossler2011}.\\

In an elementary single-particle description \cite{Beenakker1991,Berggren1998}, the size of the energy gaps for the even-numbered filling factors can be read from the finite bias diamonds (see Fig 3a), dashed lines). The extracted gaps correspond to the expected values which can be calculated according to $\Delta E=\hbar\omega_c$. The finite bias measurements in Figs. 3 a) and b) are rather different in their details. The reason lies in the fact that in Fig. 3a) we look at a QPC which is formed between the tip and the left gate, whereas in Fig. 3b) we measured the QPC formed between the two gates. Furthermore the two measurement were made in two different cooldowns. \\

\begin{figure}
\centering
\includegraphics[width=\columnwidth]{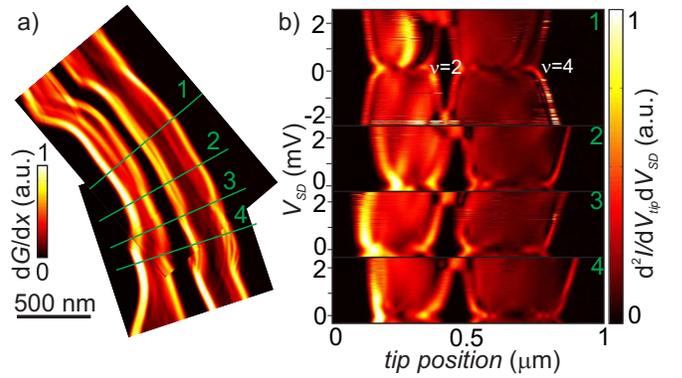}
\caption{a) Zoom on the feature close to line number 1 in Fig. 1 b). The structures are rugged on a scale of several 100 nm. To examine the position dependence of the energy gaps linescans were taken along lines 1 - 4. b) From the closing point of the black diamonds the values for the energy gaps can be deduced. They are the same for every line, thus the energy gaps are not position dependent, despite the irregularity of the observed feature.}
\label{Gaps}
\end{figure}

The irregularity of the structures in Fig. 2 was attributed to edge-roughness, which is accompanied by a complicated and irregular local electrostatic landscape. Yet one could imagine, that due to different local properties of the 2DEG also the local energy gaps could be different. To check this, different lines across the ring shaped feature were examined (as shown in Fig. 4a)) and the obtained values for the energy gaps, as they can be read from Fig. 4 b) for filling factor 2 were compared. The energy gaps are found to be position independent within experimental accuracy. 

\section{Simulations}
Paradiso et al measured the width of a conducting stripe in scanning gate images and identified this with the compressible edge state region in the potential landscape \cite{Paradiso2010,Paradiso2012}. In order to obtain the width of compressible and incompressible regions in real space one needs to carefully consider the gate as well as tip-induced potentials which are known only with limited accuracy. We start from a simple model capturing the essentials of the involved potentials and then calculate the regions of constant conductance in scanning gate images.\\

\begin{figure}
\includegraphics[width=0.8\columnwidth]{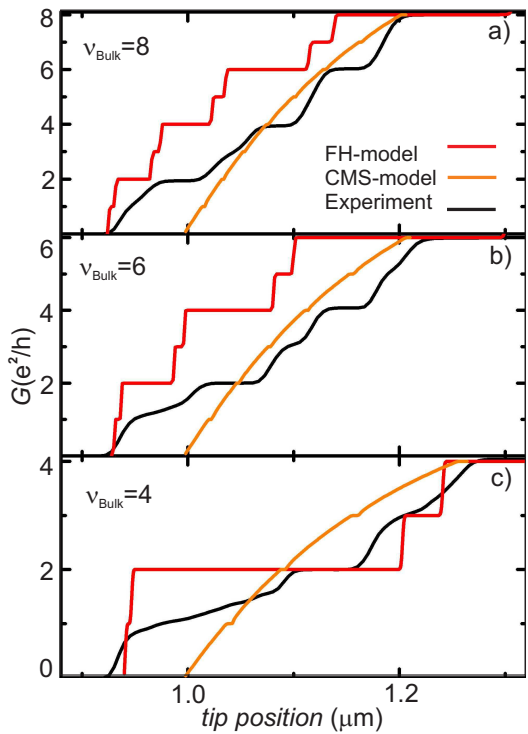}
\caption{Results for bulk filling factors 8, 6 and 4, are shown in a), b) and c). The black curve corresponds to measured traces at the position of the green line in fig. 1a). Red and orange lines were obtained performing calculations according to \cite{Chklovskii1993} and \cite{Fertig1987,Buttiker1990} respectively.}
\label{figure4}
\end{figure}

We performed calculations using the saddle point model in a magnetic field by Fertig and Halperin \cite{Fertig1987,Buttiker1990} (FH-model), where electron-electron interactions are neglected. We compare the results with the model by Chklovskii, Matveev and Shklovskii (CMS-model), where electron-electron interactions are taken into account \cite{Chklovskii1993}. For both methods the potential landscape of the QPC as influenced by the scanned tip was approximated as a superposition of a saddle point potential (QPC) \cite{Buttiker1990} and a Lorentzian potential (tip) \cite{Gildemeister2007}\\

\begin{equation}
\begin{split}
	V_{tot}=-\frac{1}{2}m^*\omega_x^2+\frac{1}{2}m^*\omega_y^2+\\
	 \frac{V_0 \gamma^2}{\gamma^2+(x-x_0)^2+(y-y_0)^2},
\end{split}
\end{equation}

where $\omega_x$ and $\omega_y$ describe the curvature of the saddle point potential in $x$- and $y$- directions as extracted from finite bias measurements at $B=0$ T \cite{Rossler2011}, $V_0$ describes the height of the Lorentzian potential which is proportional to the voltage applied to the tip, and $\gamma$ denotes the half width at half maximum of the tip-induced potential. If the tip is scanned across the saddle point, the coordinates $x_0$ and $y_0$ describe its position.\\

If the tip is moved close to the QPC, $V_{tot}$ describes a potential landscape, where saddle points which form in between the tip and the QPC walls dominate the current flow. The positions ($x_S,y_S$) of extremal points of these potentials can be found with the condition

\begin{equation}
	\frac{\partial V_{tot}(x,y)}{\partial x}=\frac{\partial V_{tot}(x,y)}{\partial y}=0 . 
\end{equation}

Expanding the total potential in a power-series around the saddle point found in this way, one obtains

\begin{equation}
\begin{split}
	V_{tot}=V_{tot}(x_S,y_S)+\\
	 \frac{1}{2}\frac{\partial^2 V_{tot}}{\partial x^2}\vert_{(x_S,y_S)} (x-x_s)^2+\\
	 \frac{1}{2}\frac{\partial^2V_{tot}}{\partial y^2}\vert_{(x_S,y_S)} (y-y_s)^2 ,
	\end{split}
\end{equation}

with the terms $\frac{\partial^2 V_{tot}}{\partial x^2}\vert_{(x_S,y_S)}<0$ and $\frac{\partial^2V_{tot}}{\partial y^2}\vert_{(x_S,y_S)}>0$ which can be identified as the shape-parameters $-m^*\omega_{xs}$ and $m^*\omega_{ys}$ of the saddle point which is formed in-between the tip and the gates. According to \cite{Fertig1987,Buttiker1990} only these shape-parameters of the potential at the saddle point and the relative position of the potential minimum towards the Fermi energy are needed to calculate the transmission. It has been calculated within the FH-model for each position of the tip on a line across the center of the QPC (green line number 2 in fig. 1b)). This model accounts for non-interacting electrons in a potential landscape, however the included spin-splitting uses a g-factor which accounts for many particle interactions, as can be estimated from finite-bias measurements in a magnetic field \cite{Rossler2011}.\\

Also the calculations which follow the CMS-model use the electrostatic model developed above to calculate the electrostatic width $2b$ of the channel at the Fermi energy $E_F=6.7$ meV as a function of tip position. This is the width of the confining parabola in $x$-direction at $E_F$. From this width the electron density in the center of the channel can be estimated to be \cite{Chklovskii1993}

\begin{equation}
	n=n_0\frac{b}{d},
\end{equation}

where $d$ is the lithographic width of the channel of $800$ nm, $n_0=1.9\times10^{11}cm^{-2}$ is taken to be the bulk density of the 2DEG as determined by QHE measurements. Using the channel width and the density in the center of the channel, the transmission can be calculated for each position of the tip across the center of the QPC (green line number 2 in fig. 1b)) according to eqs. (48), (50) and (51) in \cite{Chklovskii1993}.
If the tip is placed close to the QPC-gap it is intuitively understandable that the superposition of potentials leads to a modified transmission of current, as the minimum of the confining potential is raised relative to the Fermi energy and the width of the channel is also changed.\\

In Fig. 5. the results of the FH- and CMS-models are compared to experimental traces. In both calculations plateaus can be interpreted as the result of incompressible stripes, while the regions of finite slope represent the compressible stripes in the channel center. One can observe the trend that the calculations which follow the FH-model (red curves in Fig. 5) overestimate the plateau width and underestimate the width of the sloped regions, thus giving an exaggerated impression of the incompressible stripe width. On the other hand the CMS-model (orange curves in Fig. 5) overestimates the width of the sloped regions and greatly underestimates the plateau widths, leading to compressible stripes which are broader than observed experimentally. Our calculations are performed at zero temperature. Including the effect of finite temperature would lead to a stronger smearing of the plateaus. The experiment is found in-between the extreme results of the two models. This may not be very surprising. The FH-model is based on non-interacting electrons. Self-consistent calculations \cite{Ihnatsenka2006} indicate that edge-reconstruction due to screening cannot be neglected in the case of smooth gate-defined edges. However, the CMS-model assumes perfect metallic screening by compressible stripes, an assumption that exaggerates the influence of interactions. \\

\section{Conclusions}
In our measurements scanning gate microscopy has proven to be a powerful tool to explore the integer and fractional quantum Hall effects, adding to conventional transport experiments the possibility to image in real space. Stripes of constant conductance in the maps of conductance as a function of tip position can be identified as the incompressible regions, which form in the center of the constriction in the quantum Hall regime. Their real space and energy distribution can be studied in greater detail. Taking the very local nature of the edge channels into account it seems surprising, that finite bias spectroscopy measurements show unambiguously, that the corresponding energy gaps of incompressible regions of a particular integer filling, here filling factor 2 factor do not depend on the position of the tip in the potential landscape. Putting our results in a theoretical context we can show, that existing analytical theories qualitatively reproduce the results, but have a tendency to overestimate or underestimate screening effects. Methods like SGM which give direct insights into the real space behavior of electron transport can connect the theoretical understanding of edge-channels to the world of experiments.

\section{Acknowledgments}
The authors acknowledge the Swiss National Science Foundation, which supported this research through the National Centre of Competence in Research "Quantum Science and
Technology" and the Marie Curie Initial Training Action (ITN) Q-NET 264034. We thank R. Gaudenzi, Y. Meir, Y. Gefen, M. B\"uttiker, B. Rosenow and M. Treffkorn for fruitful discussions.

\end{document}